# Hydrogen sensing characteristics of perovskite based calcium doped BiFeO$_3$ thin films

Arindam Bala[1], S. B. Majumder[1], Moumita Dewan[1] and Ayan Roy Chaudhuri[1,2]*

1. Materials Science Centre, Indian Institute of Technology Kharagpur, 721302 Kharagpur, West Bengal, India
2. Advanced Technology Development Centre, Indian Institute of Technology Kharagpur, 721302 Kharagpur, West Bengal, India

## Abstract:

Perovskite oxide based thin film gas sensors have long been considered as potential alternatives to commonly investigated binary metal oxides based sensors. BiFeO$_3$, which is a prototype of p-type perovskite based semiconducting oxides, has recently drawn significant attention for its promising gas sensing characteristics. In the present work, the hydrogen sensing characteristics of calcium doped BiFeO$_3$ has been reported by varying the film thickness, doping concentration, operating temperature, and test gas concentration. The films were deposited on glass substrates by sol-gel route using spin coating. X-ray diffraction analyses confirmed formation of phase pure films and scanning electron microscopy confirmed their uniform and dense microstructure. The Ca-doped BiFeO$_3$ sensors exhibit higher sensitivity compared to pure BiFeO$_3$ sensors. It is reported that the film thickness and Ca doping concentration play major role to control hydrogen sensing characteristics of the deposited films. The sensor based on 15% Ca-doped BiFeO$_3$ sensor exhibited very high sensitivity (~212 % at 500 ppm H$_2$), and excellent selectivity towards hydrogen at a moderate operating temperature (~250 °C).The enhanced gas sensing response of the doped BiFeO$_3$ films has been attributed to the higher oxygen vacancy concentration induced by incorporation of aliovalent Ca$^{2+}$.

* Corresponding author: Tel: +91-3222-283978

Electronic mail: ayan@matsc.iitkgp.ac.in (Ayan Roy Chaudhuri)

# 1. Introduction:

Hydrogen ($H_2$) is the most powerful alternative clean energy source, owing to its high energy density of 120−142 MJ kg$^{-1}$, which is three times that of fossil fuels. It is commercially utilized in the fuel cells of electric vehicles.[1] Besides, hydrogen is useful also as a reducing agent in variety of chemical industries.[2, 3] However, $H_2$ which has the lowest atomic number is susceptible to leakage, has low ignition energy (0.02mJ), a wide flammable range (4−75%) and large flame propagation velocity.[4] Therefore, it is highly desirable to accurately detect hydrogen during its production, storage and use. Hydrogen gas being colourless and odourless cannot be detected by human sensory organ. It is therefore pertinent to develop highly sensitive, highly selective, and low-cost portable sensors for fast detection of hydrogen gas. One of the most promising types of such sensors is semiconducting metal oxide (SMO)-based chemoresistive gas sensors that offer high sensitivity, long-term stability, low fabrication costs, and compatibility with standard semiconductor technology. Diverse SMOs, such as ZnO, $SnO_2$, $In_2O_3$, $WO_3$, $TiO_2$, CuO, NiO, etc. have been widely investigated for hydrogen detection.[5] Especially SMO based gas sensors with reduced dimension have become increasingly important as they offer possibility of on chip integration with the control electronics for the purpose of developing portable sensors.[6] SMO thin film based hydrogen sensors on flexible substrates to improve their versatility have also been addressed.[7] However, in general there are several major issues associated with commonly used binary SMO based gas sensors which include high working temperature, poor stability, sensitivity and selectivity. In the quest of suitable alternative SMOs, various spinel and peorvskite based oxides have been explored.[8-10] Perovskite oxides are particularly attractive, because, in addition to often having high melting and/or decomposition temperatures, they can provide microstructural and morphological stability to improve reliability and long-term sensor performance. Further, presence of two differently-sized cations in the perovskite structure makes it amenable to a variety of dopant additions. This doping flexibility allows for control of the transport and catalytic properties to optimize sensor performance for particular applications. Although diverse perovskite based SMOs, especially different ferrites have been explored for their sensing behaviour towards different reducing gases,[11-14] their suitability in sensing hydrogen is not widely reported in the literature.

$BiFeO_3$ (BFO) is a very interesting multifunctional material in the perovskite ferrite family. BFO, which is conventionally known for its room temperature multiferroic properties, is a semiconducting material with a band gap of ~2.2 – 2.8 eV.[15, 16] BFO has been demonstrated as a suitable gas sensing material as early as in 1991.[17] However, surprisingly this material has not been widely explored for its gas sensing characteristics. Of late, exploration of BFO as a gas sensing material has drawn renewed interest due to its reasonably high response in sensing various gases, such as ammonia, CO, $SO_2$, LPG, ethanol, formaldehyde etc.[18-22] However, there remains a lot more room for systematic investigations to determine the suitability of BFO in gas sensing device applications. For example, tuning of resistance via suitably chosen aliovalent dopants is known to enhance the response and gas selectivity of different oxide semiconductor based sensors by electronically and chemically sensitizing the SMOs respectively. In case of BFO, p-type substitution by Ba has been found to exhibit significantly enhanced gas sensing performance for ethanol compared to undoped BFO.[14] Very recently W-doped BFO nanocrystalline thin films have been reported to exhibit encouraging sensing characteristics towards $H_2$.and $NO_2$.[23] However, to the best of our knowledge in depth investigation of hydrogen sensing properties of BFO based thin films does not exist in the literature. In this paper we report the sensing properties of phase pure and Ca-doped BFO (CBFO) thin films towards H$_2$. Ca$^{2+}$ (r$_{Ca}^{2+}$ ~1.12 Å) that replaces Bi$^{3+}$ (r$_{Bi}^{3+}$ ~ 1.17 Å) acts as a p-type substitution in BFO and increases the conductivity.[24, 25] Herein, we report the hydrogen sensing characteristics of CBFO thin films by varying the doping concentration, film thickness, H$_2$ concentration and sensor operating temperature. We demonstrate that CBFO thin films offer robust sensitivity and excellent selectivity towards H$_2$ in the moderate working temperature ranging between ~ 225-250 °C.

# Experimental:

**Materials and synthesis**

BFO and CBFO precursor for thin film deposition have been prepared by sol-gel method. The BFO sol have been prepared by dissolving bismuth nitrate pentahydrate (Bi(NO$_3$)$_3$.5H$_2$O), and iron nitrate nonahydrate (Fe(NO$_3$)$_3$.9H$_2$O) (1:1 molar ratio) in 10ml 2-methoxyethanol (2-MOE) and stirring for 120 minutes at 40˚C. In order to prepare CBFO, various amount of calcium nitrate tetrahydrate (Ca(NO$_3$)$_2$.4H$_2$O) has been added for varying the calcium content (5%-20%). 2ml Glacial acetic acid and 0.3ml polyethylene glycol have been used as a chemical additive to maintain the viscosity and stabilize the precursor sol. The concentration of the stock solutions was maintained at 0.2M. For the purpose of depositing BFO and CBFO films, the sols were drop casted and spin coated at 3000rpm for 20 s on pre-cleaned glass substrates. After each coating, the films have been inserted into a preheated tube furnace at 200˚C for 5 minutes to remove the organics and then quenched to room temperature. After the final coating (4,8,12 and 16 times and 5%,10%,15% doping concentration), the films have been annealed at 600˚C for 1 hour in the air for phase formation and crystallization. Interdigitated platinum electrodes have been deposited by a DC sputtering on the film surface to measure the resistance transient.

**Materials characterization**

Phase formation of the BFO and CBFO thin films have been investigated by X-ray diffraction (Ultima III, Rigaku Japan) in ω - 2θ geometry within 2θ range of 20˚ to 70˚ using a Cu K$_\alpha$ radiation source. The surface morphology and film thickness of the annealed films have been studied using field emission scanning electron microscopy (Merlin, Gemini 2, Germany).

**Sensor measurements**

A dynamic homemade computer controlled gas sensing measurement unit equipped with a temperature controller (6400 West Instruments.) and an electrometer (6517, Keithley instruments, USA) that acts as a source measure unit has been used to measure the hydrogen sensing response of the films. The concentration of the test gas was maintained using a mass flow controller (PR 4000, MKS Technology and Products). Details about gas sensing measurement set-up can be found elsewhere.[26] Gas sensing characteristics of the CBFO thin films with were systematically characterized by varying the film thickness, Ca-doping concentration, H$_2$ concentration and temperature. From the measured value of equilibrium resistance in air ($R_a$) and test gas ($R_g$), the response of the sensor was calculated using the following relation.

$$\text{Response (\%) }(S) = (R_g - R_a)/R_a \times 100 \quad \text{... (1)}$$

During the detection of the gases, the response and recovery times of the BFO and CBFO thin film sensors have been estimated as the times required for 90% change of their resistance from the resistance measured in air (response time, $\tau_{res}$) or the resistance measured in the test gas (recovery time $\tau_{rec}$)

## Results and Discussion:

**Crystal structure and morphology of thin films**

Figure 1 shows the grazing incidence XRD patterns of BFO and CBFO films with different Ca content. All the films exhibited a single phase perovskite structure and no impurity or secondary phase was observed. The diffraction peaks exhibited by the BFO thin film correspond to the rhombohedral structure with space group *R3c* (JCPDS card 01-070-5668). Diffraction peaks (104) and (110) are separated for pure BFO. The doublet merged to a single peak in the CBFO samples and shifted to large 2θ value. Also the (006) peak of BFO got suppressed in the CBFO samples. The change in the XRD pattern in BFO thin films due to Ca-doping could be attributed to the compressive lattice distortion in BFO due to the partial substitution of $Bi^{3+}$ by $Ca^{2+}$ having smaller ionic size.[27-29] Previously it has been demonstrated that doping BFO with $Ca^{2+}$ also changes the crystal structure from distorted rhombohedral to orthorhombic for large doping concentration (≥ 20%).[28] The Goldschmidt tolerance factor of perovskite BFO (0.84036) reduces by Ca-doping (0.83966 for 5%-0.83759 for 20%). In case of CBFO in order to accommodate the smaller size $Ca^{2+}$ in the $Bi^{3+}$ site, buckling of oxygen octahedra takes place giving rise to a lattice distortion which eventually suppresses the rhombohedral phase of BFO.[27]

Figure 2 (a and b) exhibits the surface morphologies of a 16 times coated representative CBFO (10% Ca) and BFO films respectively. Thickness of the 16 times coated samples have been estimated to be ~467 nm, as shown in the cross sectional SEM image of a representative sample in figure 2c. Thickness of the 4, 8 and 12 times coated samples have been found to be ~ 250 nm, ~ 366 nm and ~ 417 nm respectively (not shown). As evident from figure 2 (a and b), particle size in the films reduces on incorporation of Ca dopant. Figure 3 shows the surface morphologies of representative CBFO samples with (a) 5% (b)10% (c) 15% and (d) 20% calcium content. All the films exhibit dense and uniform microstructure. As envisaged from the fig. 4(a–d) the particle size in CBFO films reduced with increasing Ca content. The reduction in the particle size in CBFO films indicate that doping of BFO with Ca inhibits the grain growth, which could be attributed to the concentration of doping ions near the grain boundaries and reduction of the grain-boundary mobility. Inhibition of grain growth due to grain boundary aliquation in doped BFO ceramics has been reported frequently in the literature.[14, 27, 30, 31]

**Gas sensing characteristics of bismuth ferrite and calcium doped bismuth ferrite thin films:**

The hydrogen sensing characteristics of the BFO and CBFO thin films have been measured by varying the temperature over a wide range (175°C to 300°C). Figure 4 exhibits the temperature dependent response (%) of BFO and a representative CBFO (Ca~10%, denoted as CBFO10) sensor for $H_2$ concentration fixed at 500 ppm. In both the cases, the sensor response initially increased with temperature, attains a maximum and then reduced with further increase in the operating temperature. One reason of the reduced sensitivity at higher temperature could be the enhancement of desorption rates at the sensor surface. A detailed discussion of the plausible physical mechanism controlling such trend can be found elsewhere.[32] It is observed that the response of the CBFO10 sensor is higher compared to the BFO sensor, which indicates that partial substitution of Bi-site by $Ca^{2+}$ enhances the gas sensitivity of BFO effectively. Such an enhancement in gas sensing characteristics of BFO has been reported earlier for volatile organic compounds, which was attributed to increase in concentration of oxygen vacancies ($V_O^{\cdot\cdot}$s) due to incorporation of p-type Ba substitution. [14] As the gas sensing properties of thin film sensors also depend on the thickness on the sensing material, CBFO10 films with different thicknesses have been investigated for their sensing response towards $H_2$ over the same range of temperature (175-300°C) and fixed hydrogen concentration (500 ppm). All the films exhibited appreciable sensing response within the studied range of temperature. Figure 5 (a) exhibits a representative thickness dependent $H_2$ sensing response for CBFO10 films at 225 °C. It has been found that regardless of the operating temperature, the response increases with film thickness up to ~417 nm, and deteriorates on further increase of the film thickness. Similar thickness dependence of response towards reducing gases has previously been reported earlier for $SnO_2$, ZnO and $Mg_{0.5}Zn_{0.5}Fe_2O_4$ thin film sensors.[10, 33, 34] Recently, a generic theoretical model has been proposed by Ghosh *et al.* [33] to predict the variation of sensing response with thickness in chemi-resistive thin film gas sensors. In their model it was assumed that the target gas propagates through the sensing material by Knudsen diffusion, reacts with the surface oxygen following first order kinetics, and sheet conductance of the film as a function of its thickness varies non-linearly with the concentration of the target gas. The model has been validated for CO sensing properties of ZnO thin films.[33] Considering the assumptions are valid also for $H_2$ sensing characteristic of CBFO thin films with different thickness, their variation of the response with temperature have been estimated using the following equation,

$$\frac{R_g - R_a}{R_a} = \frac{a_0}{6} \times \exp\left(-\frac{E_a}{RT}\right) \left[6 + n.m_0^2 \exp\left(-\frac{E_k}{RT}\right) T^{-0.5}\right] \times \left[\frac{C_{AS}}{\cosh\left(m_0.\exp\left(-\frac{E_k}{2RT}\right) T^{-0.25}\right)}\right]^n \quad \ldots(2)$$

where, $C_{AS}$: Test gas concentration at the film surface, $a_0$: pre exponential constant, $E_a$: activation energy of the transduction process, $E_k$: activation energy of the first order kinetic reaction, T: operating temperature, n: sensitivity, R: gas constant,

$$m_0 = \left(\frac{3k_0}{4r}\right)^{0.5} \left(\frac{\pi M}{2R}\right)^{0.25} L$$, where $k_0$: rate constant, M: molecular weight of the target gas, r, average pore radius of the annealed films, L: film thickness

Figure 5(b) exhibits the experimentally obtained temperature variation of $H_2$ response for CBFO10 film thicknesses ranging between 250 nm to 467 nm (symbols). The solid lines are fitted according to equation 1. The experimental data matched quite well with the theoretical predication. The estimated activation energies and sensitivity are tabulated in table 1. As shown in the figure, response relates well with the decrease of apparent activation energy for the transduction process ($E_a$). Thus, maximum response and maximum sensitivity (~0.56) has been achieved in a 417 nm thick CBFO film with minimum $E_a$ (~ 31.59 kJ mol$^{-1}$). Similar to the earlier report,[33] $E_k$ and $m_0$ are also found to be maximum (~146 kJ mol-1 and ~2×10$^8$ respectively) for the film with maximum response. Further, to optimize the response of CBFO films, doping concentration have been varied from 5% to 20% keeping the film thickness constant at ~417 nm. Figure 6 demonstrates the variation of response (%) of 5%, 10%, 15% and 20% calcium doped BFO thin film as a function of operating temperature for a constant $H_2$ gas concentration of 500 ppm. With increasing doping concentration the response increased monotonically up to Ca-content of ~15% (S ~212% at 250 °C). However, further increasing doping concentration to 20% resulted in decline in the response.

The sensing characteristics can be explained by considering the manifestation of oxygen vacancies in the SMOs. In p-type SMOs, the $V_O^{\cdot\cdot}$ s react with atmospheric oxygen to generate holes (eqn. 2) which dominates their conduction behaviour.

$$0.5(O_2)_{gas} + V_O^{\cdot\cdot} \leftrightarrow O_O^X + 2h^{\cdot} \ldots (2)$$

Incorporation of Ca in the BFO thin films leads to an enhancement in the quantity of $V_O^{\cdot\cdot}$ s due to substitution of $Bi^{3+}$ by $Ca^{2+}$.[31] As a consequence, Ca-doping leads to higher conductivity in the CBFO samples compared to BFO.[25] Gas sensing operation of BFO or CBFO layers is based on the adsorption of neutral oxygen molecules from the atmosphere, which gets ionized to $O_2^-$, $O^-$ or $O^{2-}$ ions by attracting electrons from the semiconductor valence band and reduces resistance of the sensing layer. At the optimum operating temperature of CBFO sensors (~250 °C in the present study), the $O^-$ species can be expected to dominate over any other species.[35] On exposing the CBFO sensor to $H_2$, the oxygen adsorbates are removed by the following reduction reaction as represented in equation 3.

$$O_2 + 2e^- \rightarrow 2O^-$$

$$H_2 + O^- \rightarrow H_2O + e^- \quad ...(3)$$

This process returns the trapped electrons in the valence band and results in a recombination of electrons and holes thereby increasing the layer resistance. Since more oxygen gas can combine with the surface defects in CBFO with increasing Ca-content, the sensing response improves. Similar improvement in the gas sensing characteristics have also been reported for Ba- doped and W-doped BFO sensors.[14, 23] Increase of Ca-doping concentration has been reported to monotonically augment the $V_O$s in BFO, which leads to lowering of band gap and higher conductivity.[31] However, at the same time Ca-doping also reduces the grain size of BFO leading to a dense microstructure (Figure 3). The observed reduction in the $H_2$ sensing response of CBFO thin films on increasing the Ca-content beyond 15% could be attributed to their dense microstructure.

CBFO15 film with a thickness of ~417 nm has been investigated for response and recovery resistance transients under different $H_2$ concentration. Figure 7 exhibits the $H_2$ concentration dependent (20-500 ppm) response recovery characteristics of the CBFO15 sensor at the optimized temperature of highest response (~250°C). The sensor exhibited appreciable response (33%) even for $H_2$ concentration as low as 20 ppm and the response increased dramatically with increasing $H_2$ concentration. As shown in figure 7 for the entire concentration range of $H_2$ studied, only a nominal drift in the base resistance has been observed after recovery. This indicates a reversible sensing characteristic of the CBFO sensor. The CBFO15 sensor exhibited moderate response and recovery times under different concentration of $H_2$. For example, the response and recovery curve of CBFO15 for 500 ppm $H_2$ have been found to be 90 s and 240 s respectively.

In order to address the selectivity of CBFO thin films, sensing characteristic of the 417 nm, CBFO15 film has been measured also for carbon monoxide (CO) and methane gas($CH_4$). Figure 8 compares the response (%) of the CBFO15 sensor under 500 ppm CO, $CH_4$, $H_2$. At high temperature (> 250°C) the sensor suffers from cross sensitivity (for CO and $H_2$). However, for working temperature $\leq$ 250°C, CBFO15 exhibits superior sensitivity towards hydrogen compared to CO, and $CH_4$.

In Table 2, hydrogen sensing characteristics (e.g. response, operating temperature, response and recovery time, test gas concentration) of the present CBFO15 sensor have been compared with those reported for different other SMO thin film based sensors reported earlier. Inspecting the table, it is clear that $H_2$ sensing characteristics of CBFO thin film is superior compared to many of the routinely investigated conventional oxide based sensors.

## Conclusions

The BFO and CBFO thin films have been fabricated on glass substrate by spin coating sols synthesized via wet chemical route using nitrate based precursors. The numbers of coating (4, 8, 12 and 16 times) and firing cycles have been repeated to synthesize films with varying thickness of the CBFO samples. Further, the Ca-content in CBFO was also varied in the range of 5%-20%. The phase formation behavior and evolution of microstructure in the fabricated films have been investigated using x-ray diffraction pattern and scanning electron microscopy. All the samples have been found to be crystalline and a change in the crystal structure was apparent with changing the Ca- concentration. Grain size in the CBFO films was found to reduce with increasing concentration of Ca. Hydrogen sensing response of the thin films was measured by varying their operating temperature (175°C to 300°C) and concentration of $H_2$ gas (20 to 500 ppm). CBFO films exhibited superior response compared to BFO and the highest response was obtained for the sample with 15% Ca (~212 %) at a moderate operating temperature of ~250 °C. The sensor was able to detect efficiently very small concentration of $H_2$ as low as 20 ppm and found to be highly selective towards $H_2$, in comparison to CO or $CH_4$ at the operating temperature of maximum response. These results in comparison to conventional oxide thin film based sensors reported earlier confirm that the CBFO films are promising sensing materials for the development of low-cost, high-performance $H_2$ gas sensors.

**Acknowledgement**

The research work is partially supported by the research grant from SERB, Govt. of India vide letter ECR/2017/000498 dated 20-03-2018.

**Figure captions:**

Figure 1: X-ray diffractogram of pure $BiFeO_3$ and 5%, 10%, 15% and 20% calcium doped $BiFeO_3$ films.

Figure 2: Surface morphology of (a) 16 times coated 10% doped CBFO, (b) 16 times coated pure BFO films. (c) Cross sectional SE image of a representative 16 coated CBFO sample.

Figure 3: Surface morphology of 417 nm (a) 5%, (b) 10%, (c) 15% (d) 20% calcium doped BFO thin films.

Figure 4: Gas sensing response of the sensors based on pure $BiFeO_3$ and 10% calcium doped $BiFeO_3$ of films exposed to 500 ppm $H_2$. Lines are to guide the eyes.

Figure 5: (a) Variation of response (%) of 10% calcium doped $BiFeO_3$ thin films towards the detection of 500 ppm of $H_2$ gas at 225 °C as a function of film thickness. The line is to guide the eyes (b) Experimentally obtained temperature variation of $H_2$ response (500ppm) for film thickness (250 nm to 467 nm). The solid lines are fitted non-linearly according to equation (1)

Figure 6: Sensing response (%) of 5%, 10%, 15% and 20% calcium doped $BiFeO_3$ thin films (~ 417 nm film thickness) towards the detection of 500 ppm of $H_2$ gas as a function of operating temperature. Lines are to guide the eyes.

Figure 7: Resistance transient of 417 nm, 15% calcium doped $BiFeO_3$ film toward the detection of 20-500 ppm of $H_2$ gas measured at 250˚C.

Figure 8: Cross sensitivity of 417 nm, 15% calcium doped $BiFeO_3$ thin film in the temperature range 225-300 °C.

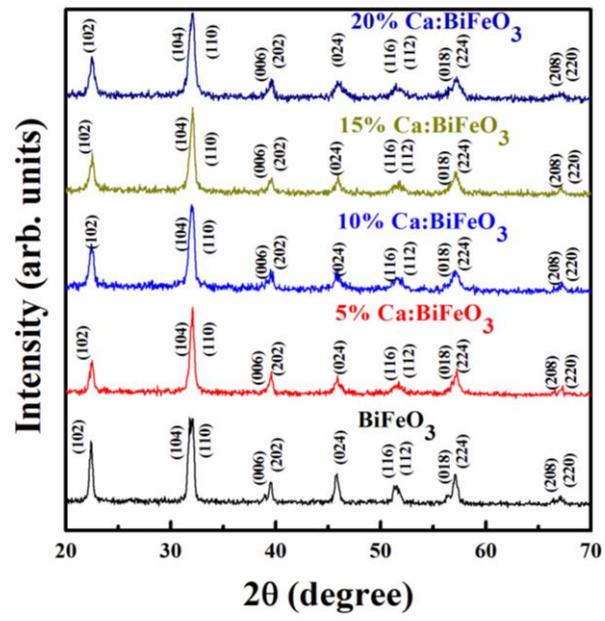

Figure 1

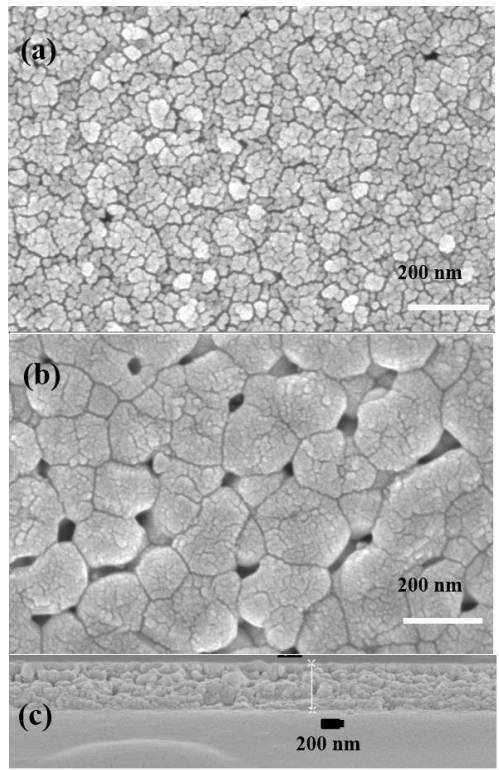

Figure 2

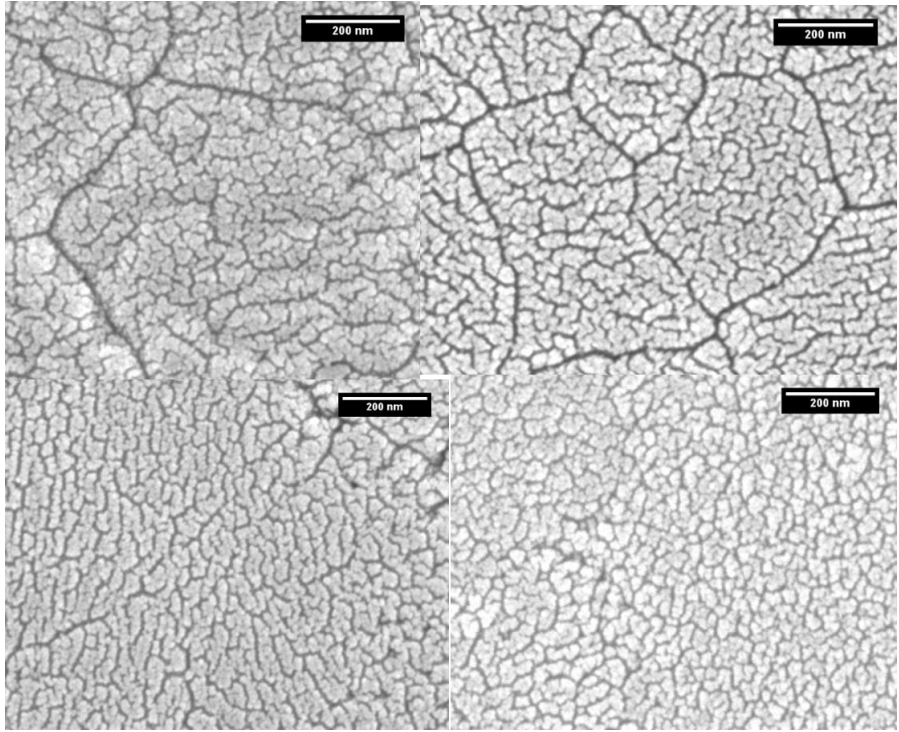

Figure 3

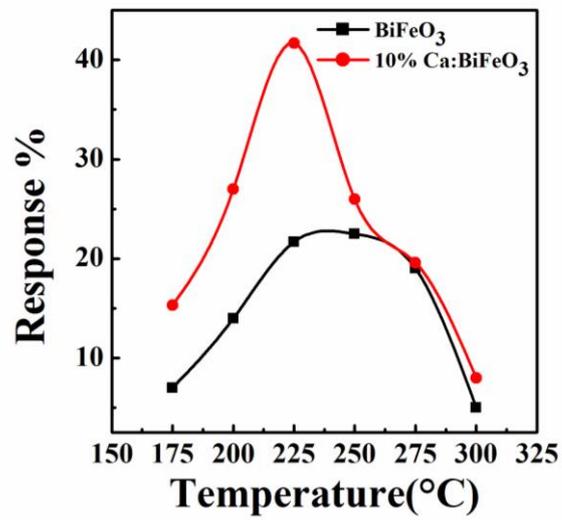

Figure 4

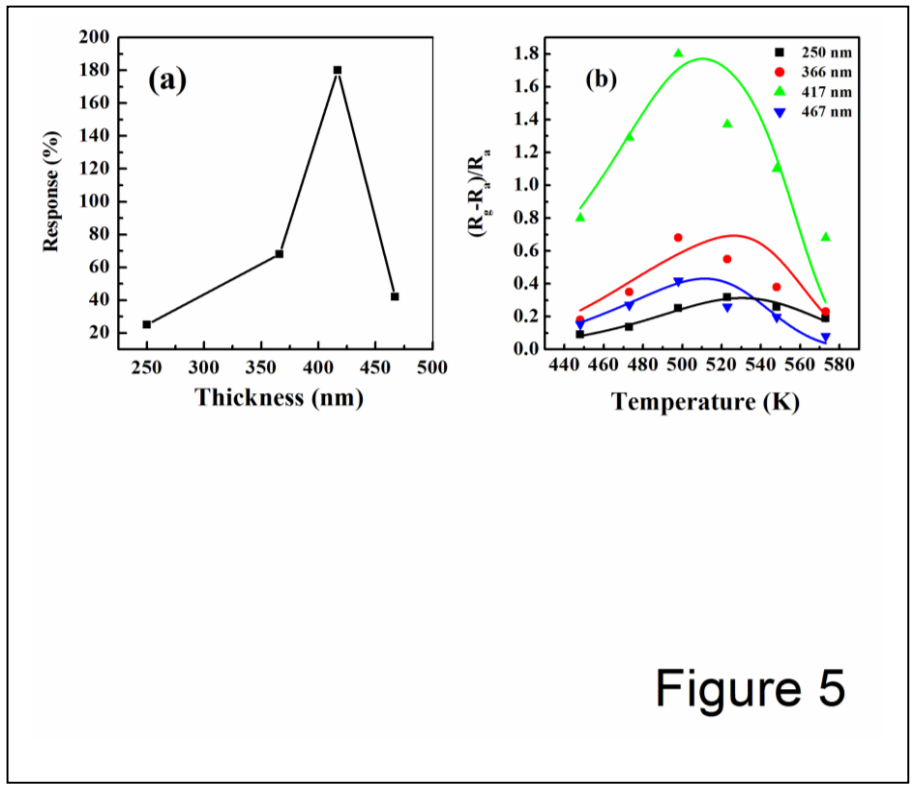

Figure 5

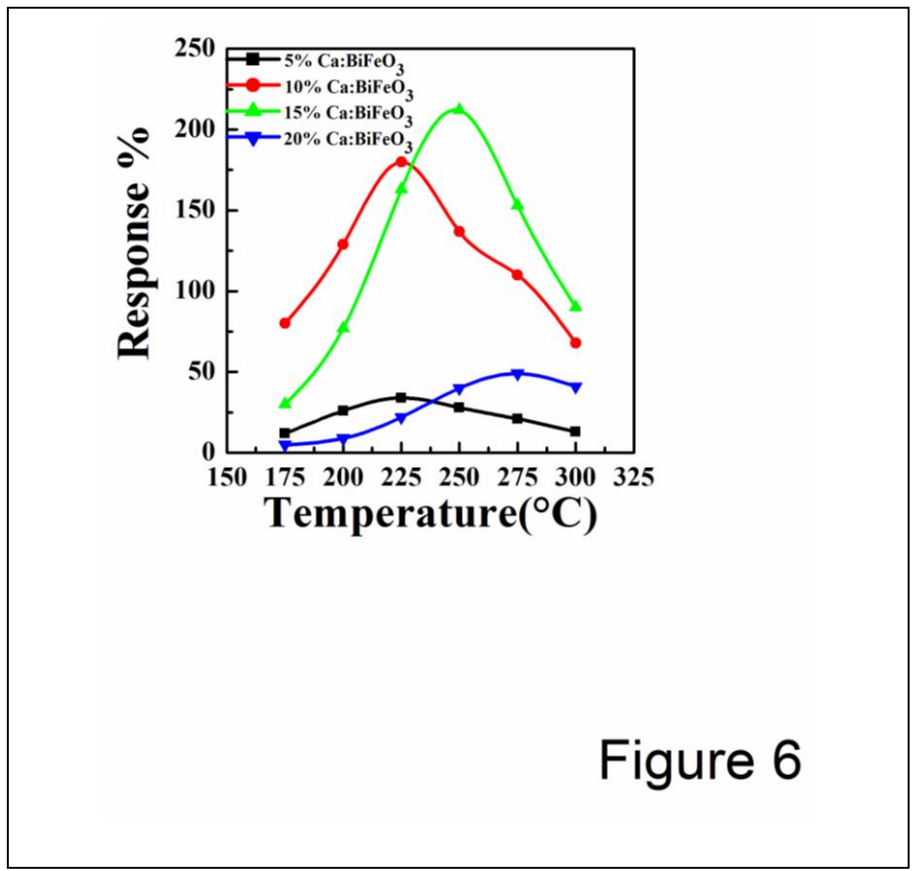

Figure 6

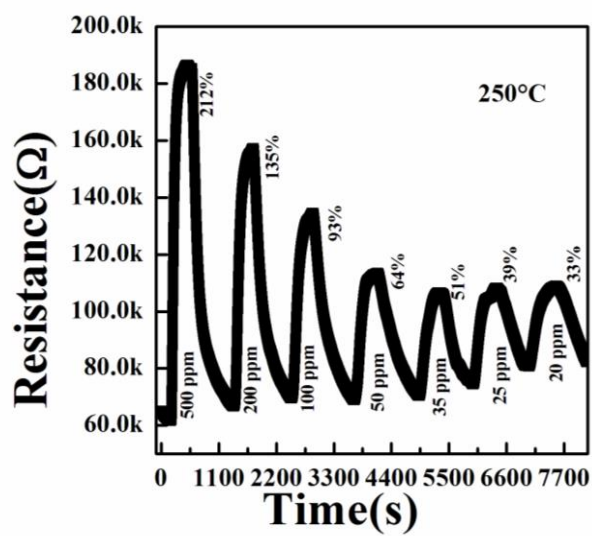

Figure 7

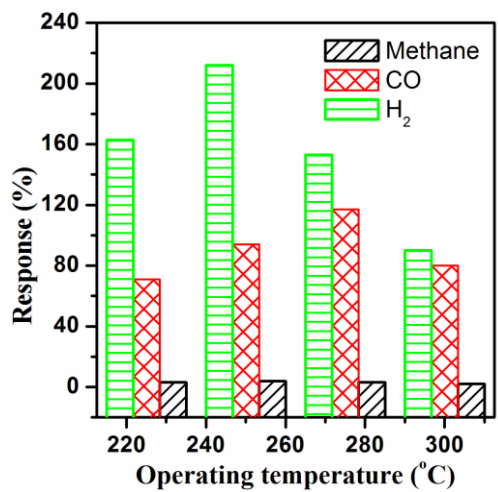

Figure 8